\documentclass[12pt]{article}
\usepackage{amssymb}
\usepackage{latexsym}
\usepackage{amsmath}
\usepackage{graphicx}

\usepackage{esdiff}
\usepackage{mathtools}

\def\vp{\varphi}
\def\ve{\varepsilon}
\def\al{\alpha}
\def\nb{\nabla}
\def\th{\theta}
\def\e{\eta}
\def\l{\lambda}
\def\k{\kappa}
\def\L{\Lambda}
\begin{document}
\title{Nonlinearly charged black hole in the theory with nonminimal derivative coupling and its thermodynamics}
\author{Mykola M. Stetsko \\{\small E-mail: mstetsko@gmail.com}\\{\small Department for Theoretical Physics, Ivan Franko National University of Lviv,}\\{\small 12 Drahomanov Str., Lviv, UA-79005, Ukraine}}

\maketitle
\abstract{ We have obtained an exact solution for a static black hole in the theory with nonminimal derivative coupling of a scalar field and gravity with a power-law Maxwell field minimally coupled to gravity. Supposing that the black hole might be topological we examined the obtained solution. We also investigated black hole thermodynamics using Wald procedure for calculation of entropy of the black hole.}

\section{Introduction}
General Relativity is extremely successful theory which allowed to explain numerous amount of facts starting from the planetary motion (perihelion of Mercury) and up to the evolution of the Universe and up to these days is in perfect agreement with high precision experiments \cite{Will_LRR2014,Berti_CQG2015}. But nonetheless there are several important issues which still remain unsolved.  Among them we should point out such problems as origin of curvature singularities, cosmological constant problem, dark energy/dark matter issue, higher order curvature corrections which give rise to the necessity of modification of General Relativity \cite{Clifton_PhysRept2012}. There are different approaches which give recipes to solve one or several of the mentioned above puzzles  and it is worth being noted that the so called scalar-tensor (or scalar-vector-tensor) theories are among the most preferable here because of their relative simplicity and minimal modification of the basic principles of General Relativity. The quite general form of the scalar-tensor theories was proposed more than for decades ago by Horndeski \cite{Horndeski_IJTP74}. Horndeski theory has not been popular among the GR community for years until it has become understood that this theory has direct relations with so called Galileon theories which is a sort of a scalar-tensor theory with specific Galilean symmetry of a scalar field \cite{Deffayet_PRD09,Kobayashi_PTEP11}. The other approach which leads to Horndeski Gravity is a dimensional reduction in the Lovelock theories \cite{Charmousis_LNP15} and we can conclude that they appear on the stage after application of  the standard technique of String Theory. The one of the most important features of Horndeski theory is the fact that it gives the equations of motion for scalar as well as the gravitational field just of the second order and it allows to avoid the existence of ghosts and instabilities typical for the theories with higher derivatives.

It should be pointed out that in its general form Horndeski theory is quite complicated and the equations one obtains are not easily tractable. That is the main reason why some particular cases of general Horndeski Gravity are undergone to deep investigation. One of the most interesting cases is the so called theories with nonminimal derivative coupling. The theory with nonminimal derivative coupling was applied to numerous problems in cosmology, namely new cosmological solution were obtained \cite{Sushkov_PRD09}, inflationary solutions and different aspects of inflation were considered \cite{Capozziello_AnnPh00,Saridakis_PRD10,Germani_PRL10,Germani_PRL11} and some other problems were also analyzed.

Apart of purely cosmological problems the investigation of black holes and other compact objects is very interesting area of research which has been developing intensively for recent years. In particular, some new solutions have been obtained in the theory with nonminimal derivative coupling and different aspects of their features have been examined \cite{Rinaldi_PRD12, Minamitsuji_PRD14,Babichev_JHEP14,Anabalon_PRD14,Kobayashi_PTEP14,
Bravo-Gaete_PRD14,Feng_JHEP15,Feng_PRD16,Sotiriou_PRL14,Maselli_PRD15,Giribet_PRD15}.

Our work is devoted to the investigation of a nonlinearly charged black hole in the theory with nonminimal derivative coupling. The work is organized in the following way. In the second section we obtain black hole's metric and investigate some of its properties. In the third section we consider thermodynamics of the  black hole. The forth section contains some conclusions.
\section{Equations of motion and black hole solution}
The action we start from consists of several terms, namely standard Einstein-Hilbert term plus cosmological constant, the terms which represent the scalar field minimally and nonminimally coupled to gravity and nonlinear Maxwell field term. So, it can be written in the form:
\begin{equation}\label{action}
S=\int d^{n+1}x\sqrt{-g}\left( R-2\Lambda-\frac{1}{2}\left(\alpha g^{\mu\nu}-\eta G^{\mu\nu}\right)\partial_{\mu}\vp\partial_{\nu}\vp +\left(-F_{\mu\nu}F^{\mu\nu}\right)^p\right)
\end{equation}
where $g_{\mu\nu}$  denotes the metric tensor, $g=det{g_{\mu\nu}}$ is the determinant of the metric, $G_{\mu\nu}$ and $R$ are the Einstein tensor and Ricci scalar correspondingly, $\L$ is the cosmological constant and $\vp$ is the scalar field, nonminimally coupled to gravity and finally $F_{\mu\nu}$ denotes the Maxwell field tensor.

The equations of motion that we derive from the action take the form as follows:
\begin{equation}\label{eom}
G_{\mu\nu}+\Lambda g_{\mu\nu}=\frac{1}{2}(\alpha T^{(1)}_{\mu\nu}+\eta T^{(2)}_{\mu\nu})+T^{(3)}_{\mu\nu}
\end{equation}
and here we use the following notations
\begin{equation}\label{scal_min}
T^{(1)}_{\mu\nu}=\nb_{\mu}\vp\nb_{\nu}\vp-\frac{1}{2}g_{\mu\nu}\nb^{\lambda}\vp\nb_{\lambda}\vp,
\end{equation}
\begin{eqnarray}\label{scal_nm}
\nonumber T^{(2)}_{\mu\nu}=\frac{1}{2}\nb_{\mu}\vp\nb_{\nu}\vp R-2\nb^{\lambda}\vp\nb_{\nu}\vp R_{\lambda\mu}+\frac{1}{2}\nb^{\lambda}\vp\nb_{\lambda}\vp G_{\mu\nu}-g_{\mu\nu}\left(-\frac{1}{2}\nb_{\lambda}\nb_{\kappa}\vp\nb^{\lambda}\nb^{\kappa}\vp\right.\\\left.+\frac{1}{2}(\nb^2\vp)^2-R_{\lambda\kappa}\nb^{\lambda}\vp\nb^{\kappa}\vp\right)
-\nb_{\mu}\nb^{\lambda}\vp\nb_{\nu}\nb_{\lambda}\vp+
\nb_{\mu}\nb_{\nu}\vp\nb^2\vp-R_{\lambda\mu\kappa\nu}\nb^{\lambda}\vp\nb^{\kappa}\vp
\end{eqnarray}
\begin{equation}\label{max_tr_nlin}
T^{(3)}_{\mu\nu}=\frac{g_{\mu\nu}}{2}\left(-F_{\l\k}F^{\l\k}\right)^p+2p\left(-F_{\l\k}F^{\l\k}\right)^{p-1}F_{\mu\rho}{F_{\nu}}^{\rho}.
\end{equation}
And for the scalar and electromagnetic fields we can write:
\begin{equation}\label{scal_f_eq}
(\alpha g_{\mu\nu}-\eta G_{\mu\nu})\nb^{\mu}\nb^{\nu}\vp=0.
\end{equation}
\begin{equation}\label{Maxwell_eq}
\nb_{\mu}\left((-F_{\l\k}F^{\l\k})^{p-1}F^{\mu\nu})\right)=0.
\end{equation}
The metric is supposed to takes the form:
\begin{equation}\label{metric}
ds^2=-U(r)dt^2+W(r)dr^2+r^2d\Omega^{2(\ve)}_{(n-1)},
\end{equation}
where $d\Omega^{2(\ve)}_{n-1}$ is the line element of a $n-1$-dimensional hypersurface of a constant curvature and it takes the form: 
\begin{eqnarray}
d\Omega^{2(\ve)}_{(n-1)}=
\begin{cases}
d\th^2+\sin^2{\th}d\Omega^2_{(n-2)}, \quad \ve=1,\\
d\th^2+{\th}^2 d\Omega^2_{(n-2)},\quad \ve=0,\\
d\th^2+\sinh^2{\th}d\Omega^2_{(n-2)},\quad \ve=-1,
\end{cases}
\end{eqnarray}
so topological black hole solution is investigated here.

The electromagnetic potential is assumed to take the only scalar component ($A=A_0(r)dt$) and as a result we obtain:
\begin{equation}\label{EM_field}
F_{rt}=\frac{q}{r^{(n-1)/(2p-1)}}\sqrt{UW},
\end{equation}
where $q$ is an integration constant, related to the charge of the black hole.

The situation with the scalar filed $\vp$ is a bit more subtle and it is difficult to work with the most general relation for the derivative $\frac{\partial \vp}{\partial r}\equiv\vp'$ which can be derived from the equation (\ref{scal_f_eq}) that is why we consider here a particular case for the scalar field which can be obtained if we impose the following condition:
\begin{equation}\label{scal_field_cond}
\al g_{rr}-\e G_{rr}=0.
\end{equation}
It should be noted that similar requirement was utilized in other works, where nonminimally coupled theory was investigated (see for example \cite{Minamitsuji_PRD14,Feng_JHEP15}).
Taking into account the relations (\ref{EM_field}) and (\ref{scal_field_cond}) we can write:
\begin{equation}\label{UW_prod}
UW=\frac{\left((\al-\L\e)r^2+\ve\e(n-1)(n-2)-2^{p-1}\e(2p-1)q^{2p}r^{2(1-p(n-1)/(2p-1))}\right)^2}{(2\al r^2+\ve\e(n-1)(n-2))^2}
\end{equation}
 \begin{eqnarray}\label{U_int}
\nonumber U(r)=\ve-\frac{\mu}{r^{n-2}}-\frac{2\L}{n(n-1)}r^{2}-2^p\frac{(2p-1)^2q^{2p}}{(n-1)(2p-n)}r^{2\left(1-\frac{p(n-1)}{2p-1}\right)}+\\\nonumber\frac{(\al+\L\e)^2}{2\al\e(n-1)r^{n-2}}\int\frac{r^{n+1}}{r^2+d^2}dr\\
+2^{p-1}\frac{(2p-1)(\al+\L\e)q^{2p}}{\al(n-1)r^{n-2}}\int\frac{r^{n+1-\frac{2p(n-1)}{2p-1}}}{r^2+d^2}dr+2^{2p-3}\frac{(2p-1)^2\e q^{4p}}{\al(n-1)r^{n-2}}\int\frac{r^{n+1-\frac{4p(n-1)}{2p-1}}}{r^2+d^2}dr
 \end{eqnarray}
 and here $d^2=\ve\e(n-1)(n-2)$. It is worth noting that the result of integration in (\ref{U_int}) depends on the dimension of space $n$, the character of exponents in the last two integrals (whether they are integer or not) and the parameter $\ve$. If $\ve=0$ we arrive at the relation:
 \begin{eqnarray}\label{flat_metr}
\nonumber U(r)=-\frac{\mu}{r^{n-2}}+\frac{(\al-\L\e)^2}{2\al\e n(n-1)}r^2-2^{p-1}\frac{(\al-\L\e)(2p-1)^2q^{2p}}{\al(n-1)(2p-n)}r^{2\left(1-\frac{p(n-1)}{2p-1}\right)}\\+2^{2p-3}\frac{(2p-1)^3\e q^{4p}}{\al(n-1)(4p-n-2pn)}r^{2\left(1-\frac{2p(n-1)}{2p-1}\right)}
\end{eqnarray}
so here we have completely power law dependence with dominating $\sim r^2$ term for large distances and $\sim r^{2(1-\frac{2p(n-1)}{2p-1})}$ for the small ones. 

For the cases $\ve=\pm 1$ the general expression for the function $U(r)$ takes the form (for odd $n$):
 \begin{eqnarray}\label{funct_odd}
\nonumber U(r)=\ve-\frac{\mu}{r^{n-2}}-\frac{2\L}{n(n-1)}r^2+\frac{(\al+\L\e)^2}{2\al\e(n-1)}\left[(-1)^{(n+1)/2}\frac{d^n}{r^{n-2}}\arctan\left({\frac{r}{d}}\right)+\sum^{(n-1)/2}_{j=0}(-1)^jd^{2j}\frac{r^{2(1-j)}}{n-2j}\right]\\\nonumber-2^p\frac{(2p-1)^2q^{2p}}{(2p-n)(n-1)}r^{\frac{2(3p-pn-1)}{2p-1}}+2^{p-1}\frac{(2p-1)^2(\al+\L\e)q^{2p}}{\al(n-1)(2p-n)}r^{2\left(1-\frac{p(n-1)}{2p-1}\right)}\times\\\nonumber{_{2}F_{1}}\left(1,\frac{p(n-1)}{2p-1}-\frac{n}{2};\frac{p(n-1)}{2p-1}-\frac{n}{2}+1;-\frac{d^2}{r^2} \right)+2^{2p-3}\frac{(2p-1)^3\e q^{4p}}{\al(n-1)(4p-2pn-n)}r^{2\left(1-\frac{2p(n-1)}{2p-1}\right)}\times\\{_{2}F_{1}}\left(1,\frac{2p(n-1)}{2p-1}-\frac{n}{2};\frac{2p(n-1)}{2p-1}-\frac{n}{2}+1;-\frac{d^2}{r^2} \right). 
 \end{eqnarray}
For even $n$ we arrive at the following expression:
\begin{eqnarray}\label{funct_even}
\nonumber U(r)=\ve-\frac{\mu}{r^{n-2}}-\frac{2\L}{n(n-1)}r^2+\frac{(\al+\L\e)^2}{2\al\e(n-1)}\left[(-1)^{n/2}\frac{d^n}{2r^{n-2}}\ln\left(\frac{r^2}{d^2}+1\right)+\sum^{n/2-1}_{j=0}(-1)^jd^{2j}\frac{r^{2(1-j)}}{n-2j}\right]\\\nonumber-2^p\frac{(2p-1)^2q^{2p}}{(2p-n)(n-1)}r^{\frac{2(3p-pn-1)}{2p-1}}+2^{p-1}\frac{(2p-1)^2(\al+\L\e)q^{2p}}{\al(n-1)(2p-n)}r^{2\left(1-\frac{p(n-1)}{2p-1}\right)}\times\\\nonumber{_{2}F_{1}}\left(1,\frac{p(n-1)}{2p-1}-\frac{n}{2};\frac{p(n-1)}{2p-1}-\frac{n}{2}+1;-\frac{d^2}{r^2} \right)+2^{2p-3}\frac{(2p-1)^3\e q^{4p}}{\al(n-1)(4p-2pn-n)}r^{2\left(1-\frac{2p(n-1)}{2p-1}\right)}\times\\{_{2}F_{1}}\left(1,\frac{2p(n-1)}{2p-1}-\frac{n}{2};\frac{2p(n-1)}{2p-1}-\frac{n}{2}+1;-\frac{d^2}{r^2} \right).
 \end{eqnarray}
Here $\mu$ is an integration constant related to black hole's mass (so called mass parameter). It should be noted that if the exponents in the last two integrals in (\ref{U_int}) are integer the resulting expression for the function $U(r)$ can be written in terms of elementary functions, namely it happens in linear case $p=1$ or so called conformal case $p=(n+1)/4$. The other important moment we have to point out here is the fact that it is assumed that $d^2>0$, because only this case leads to the black hole solution. If $d^2<0$ the solution of equations of motion also exists, but it cannot be treated as black hole, because an instability point for the scalar field $\vp$ appears outside the supposed horizon.

The obtained relations for the metric function $U(r)$ are not easily tractable but nevertheless there are some general features common for any $n$ and allowed values of $p$. In particular it can be shown that for large distances the leading term would be of AdS-type ($\sim r^2$) similarly as we had before for the case $\ve=0$ the same situation takes place for small distances, the leading term would be of the order$\sim r^{2(1-\frac{2p(n-1)}{2p-1})}$, so we arrive at the conclusion that the behaviour of the metric function (\ref{U_int}) for small and large distances does not depend on the type of geometry we use.

To additionally analyze the behaviour of the metric and find out what kind of singularity appears at different points (namely at black hole's horizon) one should use the Kretschmann scalar $R_{\mu\nu\k\l}R^{\mu\nu\k\l}$. It can be verified that at the horizons we have ordinary coordinate singularity as it has to be for any black hole. The behavior of the Kretschmann scalar for small $r$ ($r\rightarrow 0$) is of the order $\sim 1/r^4$, so we have true physical singularity and what should be emphasized here that this behaviour does not depend on $n$ neither $p$. At the infinity ($r\rightarrow\infty$) the Kretschmann scalar  behaves completely in the same manner as it is for Schwarzschild-AdS black hole.
\section{Black hole thermodynamics}
To obtain the thermodynamics of the black hole written above we start from the temperature which can be calculated in completely the same manner as it is for numerous black holes in different gravitational setup. So one obtains general formula for temperature as follows:
 \begin{equation}\label{BH_temp}
T=\frac{\k}{2\pi}=\frac{1}{4\pi}\frac{U\rq{}(r_+)}{\sqrt{U(r_+)W(r_+)}}
\end{equation}  
and here $r_+$ denotes the radius of event horizon of the black hole.
Taking into account the evident form of the functions $U(r)$ and $W(r)$ we obtain:
\begin{eqnarray}\label{temp_gen}
\nonumber T=\frac{1}{4\pi}\frac{2\al r^2_{+}+\ve\e(n-1)(n-2)}{(\al-\L\e)r^2_{+}+\ve\e(n-1)(n-2)-2^{p-1}(2p-1)\e q^{2p}r^{2\left(1-\frac{p(n-1)}{2p-1}\right)}_+}\left(\frac{(\al-\L\e)^2}{2(n-1)\al\e}r_{+}+\right.\\\nonumber\left.\ve\left(1-\frac{(\al+\L\e)^2}{4\al^2}\right)\frac{(n-2)}{r_+}+\frac{(\al+\L\e)^2}{2(n-1)\al\e}\left[(-1)^{\frac{(n+1-\sigma)}{2}}\frac{d^{n+1-\sigma}}{r^{n-2-\sigma}_{+}(r^2_{+}+d^2)}+\sum^{(n-1-\sigma)/2}_{j=2}(-1)^jd^{2j}r^{1-2j}_{+}\right]\right.\\\nonumber\left.-\frac{2^p(2p-1)q^{2p}}{n-1}r^{1-\frac{2p(n-1)}{2p-1}}_{+}\left[1-\frac{(\al+\L\e)}{2\al}\left({_{2}F_{1}}\left(1,\frac{p(n-1)}{2p-1}-\frac{n}{2};\frac{p(n-1)}{2p-1}-\frac{n}{2}+1;-\frac{d^2}{r^2_{+}} \right)-\right.\right.\right.\\\nonumber\left.\left.\left.\frac{2(2p-1)d^2}{(n+2p-2)r^2_{+}}{_{2}F_{1}}\left(2,\frac{p(n-1)}{2p-1}-\frac{n}{2}+1;\frac{p(n-1)}{2p-1}-\frac{n}{2}+2;-\frac{d^2}{r^2_{+}} \right)\right)\right]+\frac{2^{2p-3}(2p-1)^2\e q^{4p}}{\al(n-1)}\times\right.\\\nonumber\left.r^{1-\frac{4p(n-1)}{2p-1}}_{+}\left[{_{2}F_{1}}\left(1,\frac{2p(n-1)}{2p-1}-\frac{n}{2};\frac{2p(n-1)}{2p-1}-\frac{n}{2}+1;-\frac{d^2}{r^2_{+}} \right)-\frac{2(2p-1)d^2}{(2p+n-2)r^2_{+}}\times\right.\right.\\\left.\left.{_{2}F_{1}}\left(2,\frac{p(n-1)}{2p-1}-\frac{n}{2}+1;\frac{p(n-1)}{2p-1}-\frac{n}{2}+2;-\frac{d^2}{r^2_{+}} \right)\right]\right)
\end{eqnarray} 
and here $\sigma=0$ when $n$ is odd and $\sigma=1$ for even $n$. Analyzing the written above relation for the temperature we can conclude that for large radii of the horizon the temperature goes up almost linearly due to the presence of corresponding linear term and the fact that all the others terms are inversely proportional to $r_+$ and they go down to zero when $r_+\rightarrow\infty$. But for small radii of horizon these inversely proportional terms become dominant and the temperature might go up or down to infinity depending on the sign of corresponding factors near the leading terms. The temperature might have nonmonotonous beahaviour which tells as about the existence of a phase transition (supposedly of the Hawking-Page type due to the presence of the cosmological constant $\L$). 

To obtain black hole's entropy and write the first law of black hole thermodynamics we use the well grounded Wald formalism \cite{Wald_PRD93,Iyer_PRD94}. The relation of the cornerstone importance in Wald approach which allows to obtain the first law takes the form:
\begin{equation}
\delta {\cal H}_{\infty}=\delta {\cal H}_{+}
\end{equation} 
and here $\delta {\cal H}$ denotes the variation of the Hamiltonian of the problem we consider and this variation can be written as follows:
\begin{equation}\label{var_h}
\delta {\cal H}=\delta\int_{c}J_{(n)}-\int_{c}d\left(i_{\xi}\Theta_{(n)}\right)
=\int_{\Sigma^{n-1}}\delta Q_{(n-1)}-i_{\xi}\Theta_{(n)}
\end{equation} 
Having calculated the variation of the forms in the right hand side of the the latter equation we can write:
\begin{eqnarray}\label{diff_forms_h}
\nonumber\delta Q_{(n-1)}-i_{\xi}\Theta_{(n)}=r^{n-2}\sqrt{UW}\left(\frac{(n-1)}{W^2}\left(1+\frac{\e}{4}\frac{(\vp')^2}{W}\right)\delta W+\frac{2p(2p-1)2^{p-1}r}{(UW)^p}\times\right.\\\left.\left(\frac{\delta U}{U}+\frac{\delta W}{W}\right)(\phi')^{2p-1}\phi-\frac{4p(2p-1)2^{p-1}r}{(UW)^p}(\phi')^{2(p-1)}\phi\delta\phi'\right)\Omega_{n-1}
\end{eqnarray}
where $\delta W$ and $\delta U$ denote the variations of the metric functions, $\phi$ and $\phi'$ are the electric potential and its derivative with respect to $r$ (electric field) correspondingly and $\delta\phi'$ is the variation of the electric field (derivative of the scalar potential) and finally $\Omega_{n-1}$ denotes the the form over the $n-1$ angle variables which is integrated in the relation (\ref{var_h}).

Having calculated the variation (\ref{diff_forms_h}) at the infinity we can write:
\begin{equation}
\delta{\cal H}_{\infty}=\delta M-\Phi_{e}\delta Q_{e}
\end{equation}
and here $M$ and $Q_{e}$ denote mass and charge of the black hole respectively and $\Phi_{e}$ is the potential at the infinity. It is worth being noted that here we have used the gauge where the potential is equal to zero at the horizon (this gauge simplifies the calculation of the variation at the horizon). At the horizon we have:
\begin{equation}\label{TD_diff}
\delta {\cal H}_{+}=\frac{(n-1)\omega_{n-1}}{16\pi}U\rq{}(r_+)r^{n-2}_{+}\delta r_{+}=\sqrt{U(r_+)W(r_+)}T\delta\left(\frac{{\cal A}}{4}\right)=\left(1+\frac{\e}{4}\frac{(\vp\rq{})^2}{W}\Big|_{r_+}\right)T\delta\left(\frac{{\cal A}}{4}\right).
\end{equation}
where ${\cal A}=\omega_{n-1}r^{n-1}_+$ is the horizon area of the black hole. One can see that the written above relation does not allow us to identify entropy with the quarter of the black hole's horizon area as it is in standard General Relativity. To obtain reasonable relation for the entropy it was proposed to introduce an additional scalar ``charge'' \cite{Feng_PRD16} and its variation would appear in the first law but it would not be a conserved quantity. We introduce this ``charge'' $Q^+_{\vp}$ and corresponding conjugate potential $\Phi^+_{\vp}$ in the following manner:
\begin{equation}\label{sc_pot}
Q^{+}_{\vp}=\omega_{n-1}\sqrt{1+\frac{\e}{4}\frac{(\vp\rq{})^2}{W}\Big|_{r_+}}, \quad \Phi^{+}_{\vp}=-\frac{{\cal A}T}{2\omega_{n-1}}\sqrt{1+\frac{\e}{4}\frac{(\vp\rq{})^2}{W}\Big|_{r_+}}
\end{equation}
It should be noted that the choice of these ``charge'' and potential is not unique, but the given above form (\ref{sc_pot}) allowed us to obtain the so called Smarr relation \cite{Stetsko} for a chargeless slowly rotating black hole. In the end the entropy can be represented in the form:
\begin{equation}\label{entropy}
S=\left(1+\frac{\e}{4}\frac{(\vp\rq{})^2}{W}\Big|_{r_+}\right)\frac{{\cal A}}{4},
\end{equation}
It should be noted that the obtained relation for the entropy is to some extent solution independent (the structure of the given relation would be the same for all types of geometry). Finally one can write the first law in the form:
\begin{equation}\label{first_law}
\delta M=T\delta S+\Phi^{+}_{\vp}\delta Q^{+}_{\vp}+\Phi_{e}\delta Q_{e}.
\end{equation}
\section{Conclusions}
In our work we have obtained topological static black hole's solution in the theory with nonminimal derivative coupling of the scalar field  and gravity and with minimal coupling of nonlinear electromagnetic field and gravity. The obtained results generalize the previously obtained black hole solution with linear electromagnetic field \cite{Feng_PRD16} on the case of nonlinearity and also on the case of topological solutions. The relations for metric function $U$ are quite complicated but detailed analysis demonstrates that its behaviour at infinity is of AdS-type  due to the presence of the term $\sim r^2$ and it is singular when $r\rightarrow 0$ and this point is the only point of physical singularity.

Having used Wald procedure we have managed to define the basic thermodynamic values such as entropy and mass of the black hole and after that we have derived the first law of black hole's thermodynamics. To obtain reasonable relation for the entropy we had to introduce additional scalar ``charge'' $Q^+_{\vp}$  and its conjugate ``potential'' $\Phi^{+}_{\vp}$ which appeared in the first law. But this ``charge'' is not conserved and supposedly it does not appear in the relation of Smarr type.


\begin{thebibliography}{99}
\bibitem{Will_LRR2014} C.~M.~Will, Liv.~Rev.~Rel. {\bf 17}, 4 (2014).
\bibitem{Berti_CQG2015} E.~Berti, E.~Barausse, V.~Cardoso, {\it et al.} Class.~Quant.~Grav. {\bf 32}, 243001 (2015). 
\bibitem{Clifton_PhysRept2012} T.~Clifton, P.~G.~Ferreira, A.~Padilla, C.~Skordis, Phys.~Rept. {\bf 513}, 1 (2012).
\bibitem{Horndeski_IJTP74} G.~W.~Horndeski, Int.~Journ.~Theor.~Phys. {\bf 10}, 363, (1974).
\bibitem{Deffayet_PRD09} C.~Deffayet, S.~Deser, G.~Esposito-Farese, Phys.~Rev.~D {\bf 80}, 064015 (2009).
\bibitem{Kobayashi_PTEP11} T.~Kobayashi, M.~Yamaguchi, J.~Yokoyama, PTEP {\bf 126}, 511 (2011).
\bibitem{Charmousis_LNP15} C.~Charmousis, Lect.~Notes~Phys. {\bf 892}, 25 (2015).
\bibitem{Sushkov_PRD09} S.~V.~Sushkov, Phys.~Rev.~D {\bf 80}, 103505 (2009).
\bibitem{Capozziello_AnnPh00} S.~Capozziello, G.~Lambiase, H.~J.~Schmidt, Ann.~Phys. (Wiley) {\bf 9}, 39 (2000).
\bibitem{Saridakis_PRD10} E.~N.~Saridakis, S.~V.~Sushkov, Phys.~Rev.~D {\bf 81}, 083510 (2010).
\bibitem{Germani_PRL10} C.~Germani, A.~Kehagias, Phys.~Rev.~Lett. {\bf 105}, 011302 (2010).
\bibitem{Germani_PRL11} C.~Germani, A.~Kehagias, Phys.~Rev.~Lett. {\bf 106}, 161302 (2011).
\bibitem{Rinaldi_PRD12} M.~Rinaldi, Phys.~Rev.~D {\bf 86}, 084048 (2012).
\bibitem{Minamitsuji_PRD14} M.~Minamitsuji, Phys.~Rev.~D {\bf 89}, 064017, (2014).
\bibitem{Babichev_JHEP14} E.~Babichev, C.~Charmousis, JHEP {\bf 08}, 106 (2014).
\bibitem{Anabalon_PRD14} A.~Anabalon, A.~Cisterna, J.~Oliva, Phys.~Rev.~D {\bf 89}, 084050 (2014).
\bibitem{Kobayashi_PTEP14} T.~Kobayashi, N.~Tanahashi, Prog.~Theor.~Exp.~Phys. {\bf 2014}, 073E02 (2014).
\bibitem{Bravo-Gaete_PRD14} M.~Bravo-Gaete, M.~Hassaine, Phys.~Rev.~D {\bf 90}, 024008 (2014).
\bibitem{Feng_JHEP15} X.-H.~Feng, H.-S.~Liu, H.~Lu, C.~N.~Pope, JHEP {\bf 1511}, 176 (2015).
\bibitem{Feng_PRD16} X.-H.~Feng, H.-S.~Liu, H.~Lu, C.~N.~Pope, Phys. Rev. D {\bf 93} 044030 (2016). 
\bibitem{Sotiriou_PRL14} T.~P.~Sotiriou, S.-Y.~Zhu, Phys.~Rev.~Lett. {\bf 112}, 251102 (2014).
\bibitem{Maselli_PRD15}A.~Maselli, H.~O.~Silva, M.~Minamitsuji, E.~Berti, Phys.~Rev.~D~{\bf 92}, 104049 (2015).
\bibitem{Giribet_PRD15} G.~Giribet, M.~Tsoukalas, Phys.~Rev.~D {\bf 92}, 064027 (2015).
\bibitem{Wald_PRD93} R.~M.~Wald, Phys. Rev. D {\bf 48}, 3427 (1993).
\bibitem{Iyer_PRD94} V.~Iyer, R.~M.~Wald, Phys. Rev. D {\bf 50}, 846 (1994).
\bibitem{Stetsko} M.~M.~Stetsko, in preparation.
\end{thebibliography}
\end{document}